\documentclass[prl,aps,floats,twocolumn,prd,showpacs,nofootinbib]{revtex4}

\usepackage{amssymb}
\usepackage{amsmath}
\usepackage{natbib}
\usepackage{graphicx}
\usepackage{graphics}
\usepackage{dcolumn}
\usepackage{color}
\usepackage{rotate}
\usepackage{fancyhdr} 

\def\Ptot{P^{\rm tot}}
\def\Pn{P^{\rm n}}
\def\veck{\mathbf{k}}
\def\vecK{\mathbf{K}}
\def\vecx{\mathbf{x}}

\def\VEV#1{\left\langle #1 \right\rangle}

\begin{document}

\title{Clustering Fossils from the Early Universe}

\author{Donghui Jeong$^1$ and Marc Kamionkowski$^{1,2}$}
\affiliation{$^1$Department of Physics and Astronomy, Johns
     Hopkins University, 3400 N.\ Charles St., Baltimore, MD 21210}
\affiliation{$^2$California Institute of Technology, Mail Code 350-17,
     Pasadena, CA 91125}

\date{March 1, 2012}

\begin{abstract}
Many inflationary theories introduce new scalar, vector, or
tensor degrees of freedom that may then affect the generation
of primordial density perturbations.  Here we show how to
search a galaxy (or 21-cm) survey for the imprint of primordial
scalar, vector, and tensor fields.  These new fields induce
local departures to an otherwise statistically isotropic
two-point correlation function, or equivalently, nontrivial
four-point correlation functions (or trispectra, in Fourier
space), that can be decomposed into scalar, vector, and tensor
components.  We write down the optimal estimators for these
various components and show how the sensitivity to these modes
depends on the galaxy-survey parameters.  New probes of
parity-violating early-Universe physics are also presented.
\end{abstract}

\pacs{98.80.-k}

\maketitle

\section{Introduction}

Galaxy clustering has proven to be invaluable in
assembling our current picture of a Universe with a nearly
scale-invariant spectrum of the primordial curvature perturbations
\cite{clustering}.  The principle tool in clustering
studies has been the two-point correlation function, or in Fourier space
the power spectrum, determined under the assumption of
statistical isotropy (SI).  With the advent of new generations of
galaxy surveys, as well as longer-term prospects for measuring
the primordial mass distribution with 21-cm surveys
of the epoch of reionization \cite{21cm} and/or dark ages
\cite{arXiv:0902.0493}, it is worthwhile to think about what can
be further done with these measurements.

Many inflationary models introduce new fields that may couple to
the inflaton responsible for generating curvature perturbations.
The effects of these fields may then appear as local departures
from SI, or as non-Gaussianity, in the curvature perturbation.
For example, models with an additional scalar field introduce a
nontrivial four-point correlation function (or trispectrum, in
Fourier space) \cite{Baumann:2011nk}, which we below will
describe as local departures from statistical isotropy; apart
from this correlation, the scalar field may leave no visible
trace.   There may also be vector (spin-1) fields $V^\mu$
\cite{vectors}---or vector spacetime-metric perturbations
brought to life in alternative-gravity theories
\cite{Hellings:1973zz}---that, if coupled to the inflaton
$\varphi$ (e.g., through a term $(\partial_\mu
\varphi)(\partial_\nu \varphi) \partial^\mu V^\nu$)
may leave an imprint on the primordial mass distribution without
leaving any other observable trace.  Similar correlations with a
tensor (i.e., spin-2) field $T^{\mu\nu}$ (e.g., $(\partial_\mu
\varphi)(\partial_\nu \varphi) T^{\mu\nu}$) can be envisioned.
Even in the absence of new fields, there are tensor metric perturbations
(gravitational waves) that may be correlated with the primordial
curvature perturbation \cite{Maldacena:2002vr,Seery:2008ax}.
Tensor distortions to the two-point correlation function
(``metric shear'') may also be introduced at late times
\cite{Dodelson:2003bv,Masui:2010cz}, and late-time nonlinear
effects may induce scalar-like distortions to the two-point
function \cite{Pen:2012ft}.

Here we describe how the fossils of primordial tensor,
vector, and scalar fields are imprinted on the mass distribution
in the Universe today.  We express these relics in terms of
two-point correlations that depart locally from SI or
off-diagonal correlations of the density-field Fourier
components.  This formalism allows the
correlations to be decomposed geometrically into scalar, vector,
and tensor components.  We write down the optimal estimators for
these scalar, vector, and tensor correlations and quantify the
amplitudes that can be detected if these perturbations have (as
may be expected in inflationary models) nearly scale-invariant
spectra.

We begin with the null hypothesis that primordial density
perturbations are statistically isotropic and Gaussian.  This 
implies that the Fourier modes $\delta({\veck})$ of the
density perturbation $\delta({\vecx})$ (at some fixed time) have
covariances, $\VEV{\delta(\veck) \delta(\veck')} = V
\delta^D_{{\veck},-{\veck'}} P(k)$, 
where the Kronecker (Dirac) delta on the right-hand side is zero
unless ${\veck} = -{\veck'}$, $P(k)$ is the matter power
spectrum, and $V$ is the volume of the survey.  In other words,
the different Fourier modes of the density field are
uncorrelated under the null hypothesis.

\begin{figure*}[htbp]
\begin{center}
\includegraphics[width=6.4in]{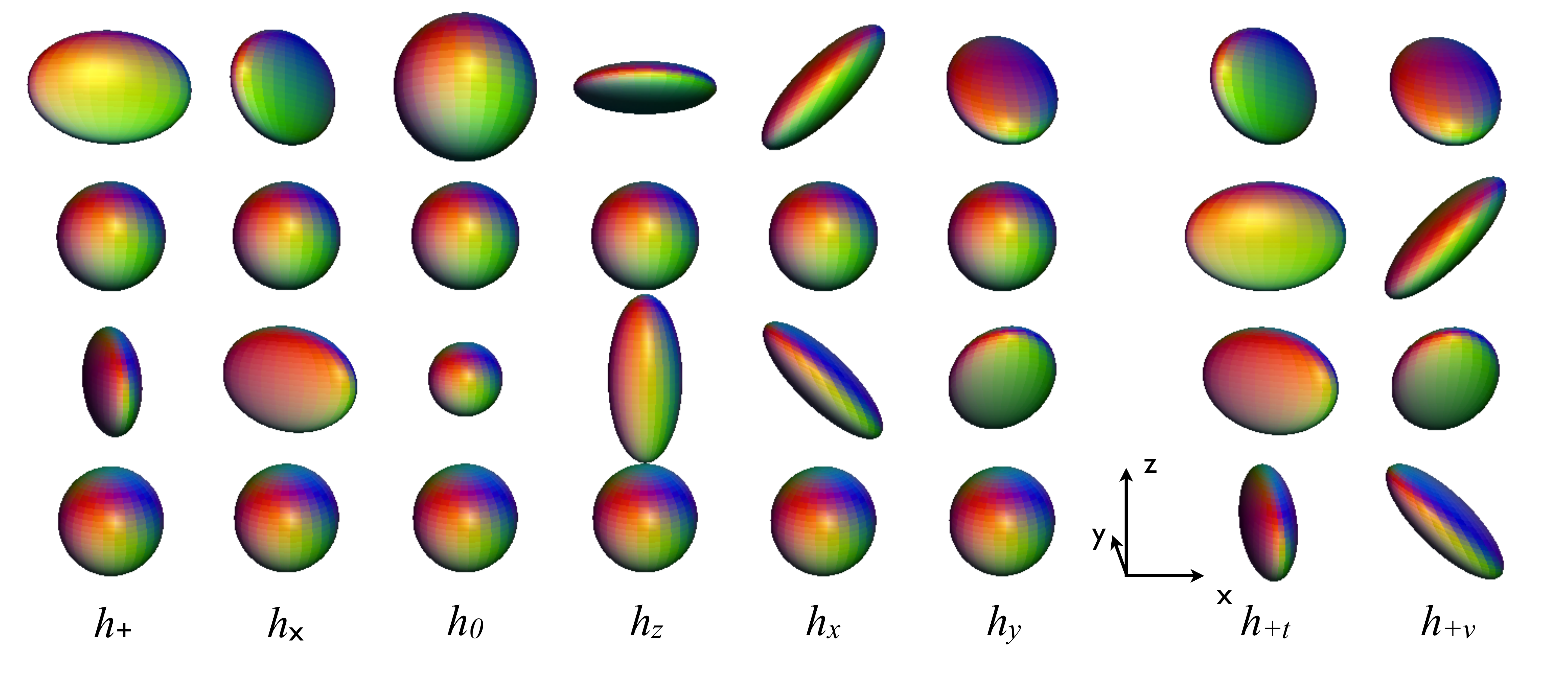}
\caption{\textit{left six}: The six possible polarizations to an otherwise
     statistically isotropic two-point correlation function.
     The first two modes are the usual transverse traceless
     tensor polarizations (gravitational waves), in which there
     are quadrupolar distortions in the plane transverse to the
     direction $\mathbf{\hat z}$ of the wave.  The next two
     are scalar and longitudinal-vector distortions,
     respectively.  The scalar mode represents an isotropic
     modulation while the longitudinal-vector mode stretches and
     compresses the correlations along $\mathbf{\hat z}$.  The
     two transverse-vector modes induce quadrupolar distortions
     in the $xz$ and $yz$ directions, respectively.
     \textit{right two}: The circular polarizations of the tensor 
		mode ($h_{+t}$) and vector mode ($h_{+v}$).
}
\label{fig:picture}
\end{center}
\end{figure*}

Coupling of the inflaton to some other field produces
non-Gaussianity in the mass distribution that appears as
off-diagonal (i.e., $\veck_1 \neq -\veck_2$) correlations of the
density-field Fourier components in the presence of a given
realization of the new field.  Global SI requires that a given
Fourier mode $h_p(\vecK)$ of wavevector $\vecK$ and polarization
$p$ (about which we will say more below) of the new
field induces a correlation,
\begin{equation}
     \left. \left< \delta(\veck_1) \delta(\veck_2) \right> 
     \right|_{h_p(\vecK)} = f_p(\veck_1,\veck_2)
     h_{p}^*(\vecK) \epsilon^p_{ij} k_{1}^i k_{2}^j
     \delta^D_{\veck_{123}},
\label{eqn:offdiagonal}
\end{equation}
where $\delta^D_{\veck_{123}}$ is shorthand for a Kronecker
delta that sets $\veck_1+\veck_2+\vecK=0$.  Note that
$h_p(\vecK)$ here are the new-field Fourier components
during inflation when their effect on primordial perturbations
is imprinted. 
The function $f_p(\veck_1,\veck_2)$ is related to the
density-density--new-field bispectrum $B_p(k_1,k_2,K)$ and
new-field power spectrum $P_p(K)$ through $B_p(k_1,k_2,K) \equiv P_{p}(K)
f_p(\veck_1,\veck_2)\epsilon_{ij}^{p}k_1^ik_2^j$.  Global SI
requires that $f_p(\veck_1,\veck_2)$ be a function only of
$k_1^2$, $k_2^2$, and $\veck_1\cdot \veck_2$.

The parameter $p$ labels the polarization state of the new 
field and $\epsilon^p_{ij}(\vecK)$ its polarization tensor, a
symmetric $3\times3$ tensor.  The most general such tensor can
be decomposed into 6 orthogonal polarization states
\cite{Eardley:1973br}, which we label $s=\{+,\times,0,z,x,y\}$,
that satisfy $\epsilon^p_{ij} \epsilon^{p',ij}=2\delta_{pp'}$.
These states can be taken to be two scalar modes
$\epsilon^0_{ij} \propto \delta_{ij}$ and $\epsilon^z_{ij}
\propto K_i K_j -K^2/3$, two vector modes $\epsilon^{x,y}_{ij}
\propto K^{(i} w^{j)}$ with $K^i w^i=0$, and two transverse
traceless modes (the ``tensor'' modes) $\epsilon_+$ and
$\epsilon_\times$.

If $\vecK$ is taken to be
in the $\mathbf{\hat z}$ direction, then the $+$ polarization of
the tensor mode has $\epsilon^+_{xx}=-\epsilon^{+}_{yy}=1$ with
all other components  zero, and the $\times$ polarization has
$\epsilon^\times_{xy}= \epsilon^{\times}_{yx} =1$ with all other
components zero.  These two tensor modes are thus characterized
by a $\cos2 \phi$ or $\sin 2\phi$ dependence,
for $\epsilon^+$ and $\epsilon^\times$, respectively, on the
azimuthal angle about the $\vecK$ direction of the tensor mode.
The first two columns in Fig.~\ref{fig:picture} show the
distortions induced to an otherwise isotropic two-point
correlation function by correlation of the density field with a
$+$ and $\times$ polarized tensor mode.  Shown there is a
quadrupolar distortion in the $x$-$y$ plane that then oscillates
in phase as we move along the direction $\mathbf{\hat z}$ of the
Fourier mode.

The first scalar mode has $\epsilon^0_{ij} =\sqrt{2/3}
\delta_{ij}$ and as shown in Fig.~\ref{fig:picture} represents
an isotropic modulation of the correlation function as we move
along the direction $\mathbf{\hat z}$ of the Fourier
wavevector. The other scalar (or longitudinal-vector) mode has
$\epsilon^z_{ij} \propto {\rm diag}(-1,-1,2)/\sqrt{3}$
which represents a stretching and compression along
$\mathbf{\hat z}$.  Both scalar modes represent local
distortions of the two-point function that have azimuthal
symmetry about $\vecK$.

Finally, the two transverse-vector modes have 
$\epsilon^x_{xz} = \epsilon^x_{zx}=1$ with all other components
zero, and  $\epsilon^y_{yz} = \epsilon^y_{zy}=1$ with all other
components zero.  These two modes represent stretching in
the $\pm xz$ and $\pm yz$ directions, respectively, as shown in
the last $h_x$ and $h_y$ columns in Fig.~\ref{fig:picture}.  These two
transverse-vector modes have $\cos\phi$ and  $\sin\phi$
dependences on the azimuthal angle $\phi$ about the direction of
the Fourier mode.

The specific functional form of $f_p(\veck_1,\veck_2)$
depends on the coupling of the new field (scalar,
vector, or tensor) to the inflaton.
Global SI requires, though, that
$f_p(\veck_1,\veck_2)$ will be the same for the two tensor
polarizations and the same for the two vector polarizations;
i.e., $f_\times(\veck_1,\veck_2)=f_+(\veck_1,\veck_2)$, and
$f_x(\veck_1,\veck_2)=f_y(\veck_1,\veck_2)$.
The same is not necessarily true for the scalar perturbations.
In fact, the polar-angle dependence that distinguishes the 0 and
$z$ polarizations can be absorbed into $f_0(\veck_1,\veck_2)$
and $f_z(\veck_1,\veck_2)$.  Thus, in practice, one can describe
the most general scalar distortions to clustering in terms of
either the 0 or the $z$ polarization by appropriate definition
of $f_0(\veck_1,\veck_2)$ or $f_z(\veck_1,\veck_2)$.  (This is
the mixing between a scalar mode and a longitudinal-vector
mode.)  We thus below merge these two polarizations into a
single polarization which we label  with a subscript $s$.

Suppose now that a correlation such as that in
Eq.~(\ref{eqn:offdiagonal}), for either a scalar, vector, or
tensor distortion, is hypothesized.  How would we go
about measuring it?  According to Eq.~(\ref{eqn:offdiagonal}),
each pair $\delta(\veck_1)$ and $\delta(\veck_2)$ of density
modes with $\vecK = \veck_1 + \veck_2$ (note that we have
re-defined the sign of $\vecK$ here) provides an estimator,
\begin{equation}
     \widehat{h_p(\vecK)} = 
     \delta(\veck_1)\delta(\veck_2) \left[
     f_p(\veck_1,\veck_2) \epsilon_{ij}^p k_{1}^ik_{2}^j \right]^{-1},
\end{equation}
for the Fourier-polarization amplitude $h_p(\vecK)$.
Since $\VEV{|\delta(\veck)|^2} = V\Ptot(k)$, where
$\Ptot(k)=P(k)+\Pn(k)$ is the measured matter power spectrum,
including the signal $P(k)$ and noise $\Pn(k)$, the variance of
this estimator is
\begin{equation}
    2V \Ptot(k_1) \Ptot(k_2) \left|
    f_p(\veck_1,\veck_2) \epsilon_{ij}^p k_{1}^i k_{2}^j \right|^{-2}.
\label{eqn:individualvariance}
\end{equation}
The minimum-variance estimator for $h_p(\vecK)$ is then
obtained by summing over all these individual
$(\veck_1,\veck_2)$ pairs with inverse-variance weighting:
\begin{align}
     \widehat{{h}_p(\vecK)} =& P_p^{n}(\vecK)
     \sum_{\veck} \frac{f_p^*(\veck,\vecK-\veck) \epsilon_{ij}^p
     k^{i} (K-k)^{j} }
     {2 V \Ptot(k) \Ptot(|\vecK-\veck|)}
		\nonumber\\
     &\hspace{2cm}\times\delta(\veck)\delta(\vecK-\veck),
\label{eqn:minimvariance}
\end{align}
where the noise power spectrum,
\begin{equation}
     P_p^{n}(K) = \left[ \sum_{\veck}
     \frac{\left|f_p(\veck,\vecK-\veck) \epsilon_{ij}^p k^i
     (K-k)^j \right|^2} {2V
     \Ptot(k) \Ptot(|\vecK-\veck|)} \right]^{-1},
\label{eqn:noisepowerspectrum}
\end{equation}
is the variance with which $\widehat{h_p(\vecK)}$ is measured.
This $P_p^n(K)$ is a function only of the
magnitude $K$ (not its orientation) as a consequence of
global SI, and for the same reason,
$P_\times(K)=P_+(K) \equiv P_t(K)$, for both the signal and
noise power spectra, and similarly $P_x(K)=P_y(K)\equiv P_v(K)$.

In general, the amplitudes $h_p(\vecK)$ arise as realizations
of random fields with power spectra $P_h(K)=A_h P_h^f(K)$, for
$h=\{s,v,t\}$, which we write in terms of amplitudes $A_h$ and
fiducial power spectra $P_h^f(K)$.  We now proceed to write the
optimal estimator for the amplitudes $A_h$.  

Each Fourier-mode estimator $\widehat{h_p(\vecK)}$ for the
appropriate polarizations ($s$ for scalar, $x$ and $y$ for
vector, and $+$ and $\times$ for tensor) provides an estimator,
\begin{equation}
     \widehat{A_h^{\vecK,p}} = \left[ P_h^f(K) \right]^{-1}
     \left[{V}^{-1} \left|\widehat{ h_p(\vecK)} \right|^2 -
     P_p^{n}(K) \right],
\label{eqn:individualestimator}
\end{equation}
for the appropriate power-spectrum amplitude.  Here we have
subtracted out the noise contribution to unbias the estimator.  If
$\widehat{h_p(\vecK)}$ is estimated from a large number of
$\delta(\veck_1)$-$\delta(\veck_2)$ pairs, then it is close to
being a Gaussian variable.  If so, then the variance of the
estimator in Eq.~(\ref{eqn:individualestimator}) is, under the
null hypothesis,
\begin{equation}
     2	\left[P_h^f(K)\right]^{-2}
	\left[ P_p^{n}(K)\right]^2.
\end{equation}
Adding the estimators from each Fourier mode with
inverse-variance weighting leads us to the optimal estimator,
\begin{equation}
     \widehat{{A}_h} = \sigma_{h}^2 \sum_{\vecK,p}
     \frac{ \left[P_h^f(K) \right]^2
     }{2\left[P_p^{n}(K) \right]^2}
     \left( V^{-1} \left|\widehat{ {h}_p(\vecK)}
     \right|^2 - P_p^{n}(K) \right), 
\label{eqn:tensorestimator}
\end{equation}
where
\begin{equation}
     \sigma_{h}^{-2} =  \sum_{\vecK,p}
     \left[P_h^f(K)\right]^2 / 2 \left[P_p^{n}(K) \right]^2.
\label{eqn:noise}
\end{equation}
For the vector-power-spectrum amplitude $\widehat{A_v}$ we sum over
$p=\{x,y\}$ and for the tensor-power-spectrum amplitude $\widehat{A_t}$
over $p=\{+,\times\}$.  Following the discussion above, the sum
on $p$ is only for $p=s$ for $\widehat{A_s}$.

The estimator in Eq.~(\ref{eqn:tensorestimator}), along with the
quadratic minimum-variance estimator in
Eq.~(\ref{eqn:minimvariance}), demonstrates that the
correlation of density perturbations with an unseen scalar,
vector, or tensor perturbation appears in the density field as a
nontrivial four-point correlation function, or trispectrum.  The
dependence of the trispectrum on the azimuthal angle about the
diagonal of the Fourier-space quadrilateral distinguishes
the shape dependences of the trispectra for scalar, vector, and
tensor modes.  To specify this trispectrum more precisely,
though, requires inclusion of the additional contribution
induced by modes $\vecK$ that involve the other
two diagonals of the quadrilateral.  Likewise, if a signal is
detected---i.e., if the null-hypothesis estimators above are
found to depart at $>3\sigma$ from the null hypothesis---then
the optimal measurement and characterization of the trispectrum
requires modification of the null-hypothesis estimators in a
manner analogous to weak-lensing estimators \cite{Kesden:2003cc}.

\begin{figure}
\centering
\includegraphics[width=0.48\textwidth]{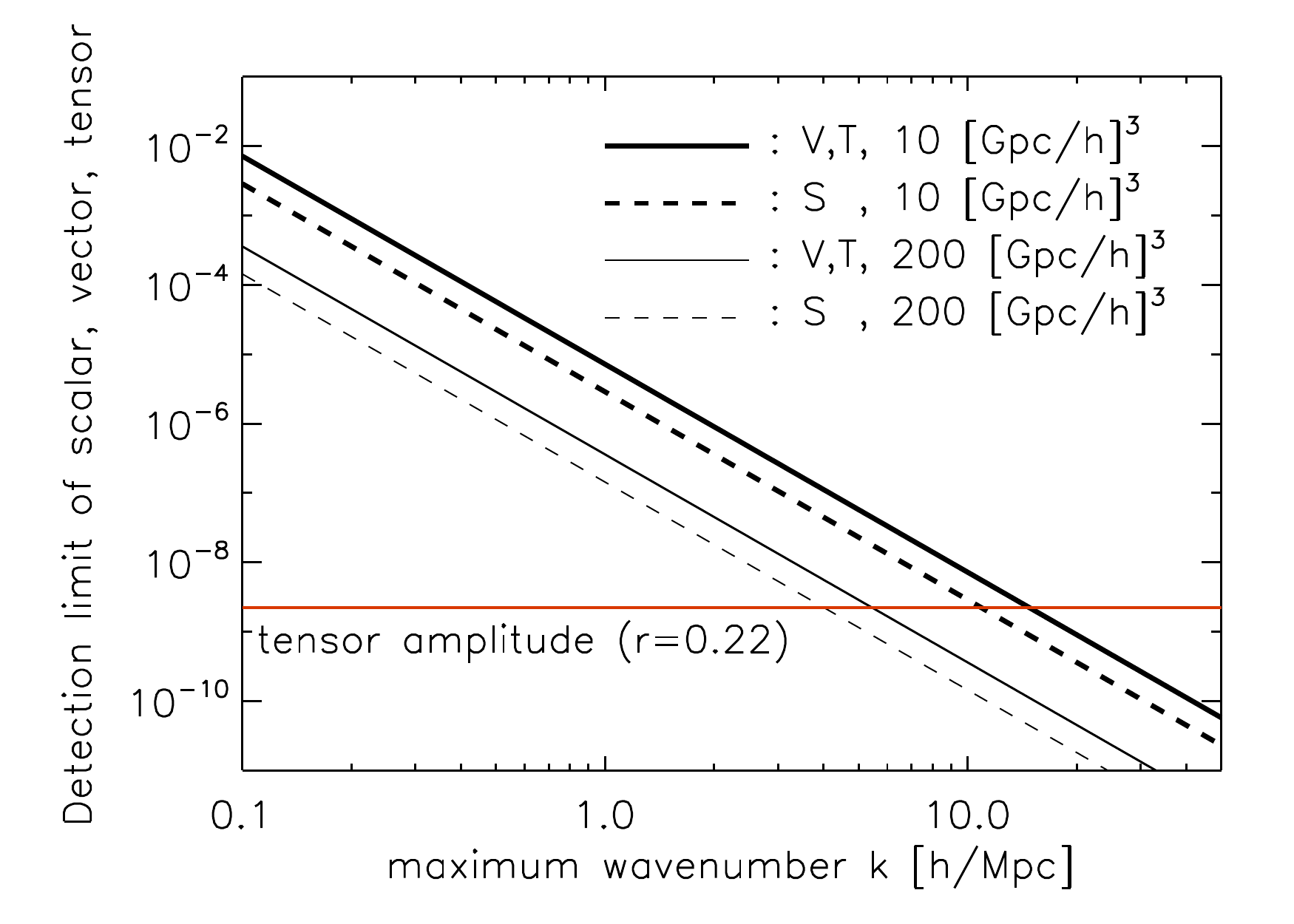}
\caption{The smallest scalar, vector, and tensor power-sepctrum
     amplitudes $A_s$, $A_v$ and $A_t$, respectively,
     detectable at the 3$\sigma$ level as a function of the maximum
     wavenumber $k_{\rm max}$ of the survey.  Shown are results
     for survey volumes of $10~[\rm{Gpc}/h]^3$ and
     $200~[\rm{Gpc}/h]^3$, or minimum wavenumbers $k_{\rm min}
     \simeq 0.001~[h/\rm{Mpc}]$ and $k_{\rm min} \simeq
     0.003~[h/\rm{Mpc}]$, respectively.
}
\label{fig:StoN}
\end{figure}

We now evaluate the smallest amplitudes $A_s$, $A_v$, and $A_t$
that can be detected with a given survey.  To do so,
we take for our fiducial models nearly scale-invariant
spectra $P_h(K)=A_h K^{n_h-3}$, with $|n_h|
\ll 1$.  We moreover take the density-density--new-field
bispectrum to be of the form in Ref.~\cite{Maldacena:2002vr}.
We then find that the integrand (using $\sum_{\veck} \to V\int
d^3k/(2\pi)^3$) in Eq.~(\ref{eqn:noisepowerspectrum}) is
dominated by the squeezed limit ($K \ll k_1\simeq k_2 $) where
$f_p(\veck_1,\veck_2) \simeq - (3/2)P(k_1) /k_1^2$.  We then
approximate $P(k)/\Ptot(k) \simeq 1$ for $k < k_{\rm max}$,
where $k_{\rm max}$ is the largest wavenumber for which the
power spectrum can be measured with high signal to noise, and
$P(k)/\Ptot(k) \simeq 0$ for $k>k_{\rm max}$.  This then yields
a noise power spectrum $P_{\{v,t\}}^{n}(K) \simeq 20\pi^2/k_{\rm
max}^3$ and $P_{s}^{n}(K) \simeq 8\pi^2/k_{\rm max}^3$.
Evaluating the integral in Eq.~(\ref{eqn:noise}), we
find the scalar, vector, and tensor amplitudes detectable at
$\gtrsim 3\sigma$ (for $n_h\simeq0$) to be
\begin{equation}
     3 \sigma_{h} \simeq 
		30\pi\sqrt{3\pi} C_h
     \left(\frac{k_{\rm max}}{k_{\rm min}} \right)^{-3} 
     \simeq 288 C_h\left(\frac{k_{\rm max}}{k_{\rm min}}
     \right)^{-3},
\label{eqn:sensivity}
\end{equation}
where $C_{\{t,v\}}=1$ and $C_s=2/5$.  The smallest detectable
power-spectra amplitudes are thus inversely proportional to the
number of Fourier modes in the survey.  We show the projected
detection sensitivities for surveys with volumes of
$200~[\rm{Gpc}/h]^3$ and $10~[\rm{Gpc}/h]^3$ in
Fig.~\ref{fig:StoN}.

For example, if we apply this estimate to a tensor field
and assume this tensor field to be primordial gravitational
waves, then a sensitivity to a tensor amplitude $A_t\simeq 2\times
10^{-9}$ near the current upper limit requires $k_{\rm
max}/k_{\rm min} \gtrsim 5200$.  Such a dynamic range is probably beyond
the reach of galaxy surveys, but it may be within reach of
the 21-cm probes of neutral hydrogen during the dark ages envisioned
in Refs.~\cite{Masui:2010cz,Loeb:2003ya}.  Of course, the signal
could be larger if the inflaton is correlated with a scalar,
vector, or tensor field that leaves no other trace.

Finally, several new tests for parity-violating early-Universe
physics can be developed from 
simple modification of the estimators above.  To do so, we
substitute the $x$ and $y$ polarizations, and $+$ and
$\times$ polarizations, with circular-polarization tensors
$\epsilon^{\pm v}_{ij} = \epsilon^x_{ij} \pm i \epsilon^y_{ij}$
and $\epsilon^{\pm t}_{ij} = \epsilon^+_{ij} \pm i
\epsilon^\times_{ij}$. The two right-most patterns shown 
in Fig.~\ref{fig:picture} are the circular polarization patterns
for tensor and vector modes.
It may then be tested whether the power
spectra for right- and left-circular polarizations are equal.
For example, chiral-gravity models \cite{Lue:1998mq} may predict
such parity-violating signatures in primordial gravitational
waves, and similar models with parity-violating vector
perturbations are easily imaginable.

Of course, ``real-world'' effects like redshift-space
distortions, biasing, and nonlinear evolution, must be taken
into account before the estimators written above can be
implemented, but there are well-developed techniques to deal
with these issues \cite{Jeong:2010}.

In summary, we have shown that the most general two-point
correlation functions for the cosmological mass distribution can
be decomposed into scalar, vector, and tensor distortions.  We
have presented straightforward recipes for measuring these
distortions.  Such effects may arise if the inflaton is coupled
to some new field during inflation.  We have avoided
discussion of specific models, but the introduction of new
fields during inflation is quite generic to inflationary
models.  We therefore advocate measurement of these
correlations with galaxy surveys, and in the future with 21-cm
surveys, as a simple and general probe of new inflationary
physics.

\bigskip

This work was supported by DoE DE-FG03-92-ER40701 and NASA
NNX12AE86G.

\end{document}